# Ab-initio-NEGF Fundamental Roadmap for Carbon-Nanotube and Two-Dimensional-Material MOSFETs at the Scaling and $V_{DD}$ Limit

Aryan Afzalian
imec, Kapeldreef 75, 3001
Leuven, Belgium
aryan.afzalian@imec.be

*Abstract*— Using accurate Hybrid-Functional DFT coupled with the Non-Equilibrium Green's function (NEGF) formalism, we explore and benchmark the fundamental scaling limits of CNT-FETs against Si and 2D-material $MoS_2$ and $HfS_2$ Nanosheets, highlighting their potential for gate length ($L$) and supply voltage ($V_{DD}$) scaling down to 5 nm and to 0.5 V, respectively. The highest drive current is achieved by CNT-FETs with sub 1.3 nm diameters down to $L$ = 9 nm and using $V_{DD}$ in the 0.45-0.5V range. Below $L$ = 9 nm, however, the $HfS_2$ NS offers the best drive and could further scale down to $L$ = 5 nm with a reduced $V_{DD}$ of 0.5 V.

## I. Introduction

Low-dimensional (LD) materials such as 2D transition-metal dichalcogenides (TMDc) or carbon nanotubes (CNTs) are explored as Si replacement for nanoscale CMOS [1], [2], [3]. In these aggressively scaled materials and devices, quantum effects such as quantum confinement (QC), tunneling and atomistic nature strongly dominate electronic properties and transport, so that efficient, ab-initio, atomistic simulation methods are best suited to model their properties [4]. Using hybrid-functional (HF) density functional theory (DFT), which is essential to accurately capture the CNT bandgap ($E_G$), coupled with NEGF transport, we investigate the scaling performance of zigzag CNT-FETs made of 4 different (n,0) chirality (n = 10, 14, 16 and 19) and benchmark them against Nanosheet (NS) transistors made of Si and 2 monolayer TMDc materials – one, $MoS_2$, is well-studied with predicted fundamental drive similar to that of Si; the other, $HfS_2$, is an emerging 2D material with an enhanced fundamental drive current [2], and for which an industry-compatible (ALD) fabrication process was recently demonstrated [5]. Our results further offer a fully ab-initio roadmap for ultra-scaled LD-material CMOS, i.e., a possible fundamental path towards gate-length, $L$, reduction down to 5 nm and supply voltage, $V_{DD}$, scaling down to 0.5 V.

## II. Methods

To compute the DFT-based Hamiltonian (H) and relaxed supercell geometries of the Si NS and CNT-FETs, we employed the DFT Package CP2K [6], using a double Zeta (DZVP) atomic-orbital basis, meta generalized gradient approximation (meta-GGA) and the PBE0-TC-LRC HF [7], with 25% of Hartree-Fock exchange and 75% PBE, in order to get an accurate $E_G$ (as fully described in [8]). For $MoS_2$ and the higher-mobility $HfS_2$ TMDc, we used QUANTUM ESPRESSO [9] and the generalized gradient approximation (GGA) with the optB86b exchange-correlation functionals [10], followed by a Wannierization step, as fully described in [2]. The resulting supercell information are then used in our quantum transport (QT) solver, ATOMOS to perform the self-consistent NEGF calculations, using a real-space H for the TMDc NS [2], [8], and a mode-space H for the Si NS and CNT-FETs [4], [11]. Electron-phonon scattering (e-ph) is included within the self-consistent Born approximation, using the deformation potential (DP) theory. We included the dominant acoustic and optical modes, as well as the polar optical mode in the case of the $HfS_2$ NS [2], [8]. For the 2D materials, we used the parameters described in [2], for the CNT the parameters from [12], while for Si, we used the parameter described in [13].

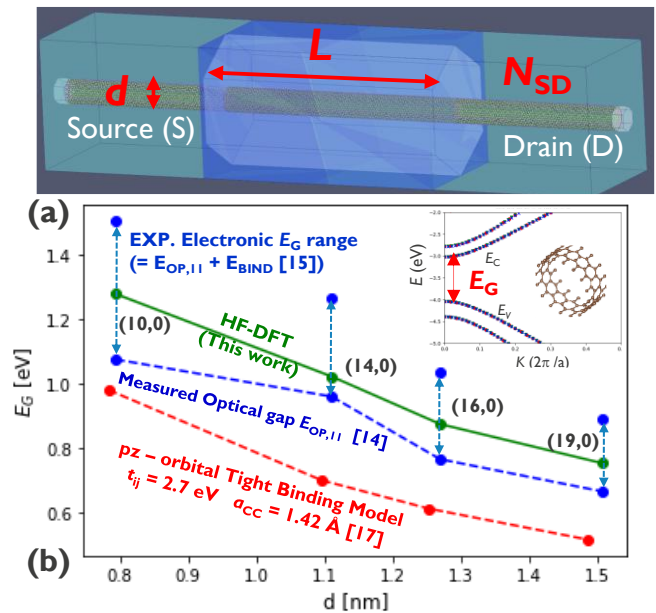

Fig. 1. (a) Atomistic view of a HF-DFT-NEGF simulated CNT-FET with gate length $L$, S&D extension doping $N_{SD}$ and diameter $d$. The cylindrical gate with an $HfO_2$ gate oxide of thickness $t_{OX}$ = 2 nm and relative permittivity $\varepsilon_R$ = 15.6, as well as spacer oxide ($\varepsilon_R$ = 4) are also shown. (b) Comparison of $E_G$ vs. $d$ relationship obtained from experiments and various models for the zigzag CNT chirality as indicated on the figure for the CNT chirality of Table I. The inset shows the atomic structure and the $EK$ dispersion of a (14,0) CNT slab computed from our real-space (red) and mode-space (blue) DFT-NEGF models. A perfect agreement can be seen.

## III. Results

Owing to their inherent nm-scale diameter, $d$, and high mobility, CNT-FETs are envisioned as high-performance (HP) nanoscale devices [3]. The desired high on current, $I_{ON}$, that could further enable a $V_{DD}$ reduction down to approximatively 0.5V, is hampered by their large band-to-band tunneling (BTBT) minimum leakage-current limit ($I_{MIN}$), arising from their low $E_G$. Fig. 1a defines the simulated CNT-FET geometry. TABLE I. and Fig. 1b show the chirality, $d$, and $E_G$ extracted from our HF-DFT simulations. In Fig. 1b, our results are further compared to experimental and tight-binding (TB) model results [14], [15]. In a CNT, the electronic $E_G$ is larger than the optically measured gap, $E_{OP,11}$, by the large exciton binding energy $E_{BIND}$ (> 100 meV), arising from many body effects. $E_{BIND}$ is, however, not directly measurable, and further sensitive to the dielectric environment and values differ between extraction models [14], [15], [16]. From the modelling side, however, TB models are known to underestimate $E_G$ by about 0.3 eV and lead to $E_G < E_{OP,11}$ ([14] and Fig. 1b), as they neglect these many-body effects. Similarly, traditional DFT methods fundamentally underestimate $E_G$. HF DFT methods, on the other hand, are well-known to provide accurate $E_G$ descriptions, including for CNTs [16], [17], at the expense of computational costs. Our HF-DFT simulated $E_G$ values are in the expected experimental windows with $E_G > E_{OP,11}$ [14], and in good agreement with those from [16] and [17] that uses HF-DFT but with different HF functionals.

TABLE I. SUMMARY OF THE CNT CHIRALITY, DIAMETERS, AND BANDGAPS, WE COMPUTED USING HF DFT (PBE0-TC-LRC).

| Chirality | $d$ (nm) | $E_G$ (eV) |
|---|---|---|
| (10,0) | 0.794 | 1.276 |
| (14,0) | 1.111 | 1.020 |
| (16,0) | 1.270 | 0.874 |
| (19,0) | 1.508 | 0.752 |

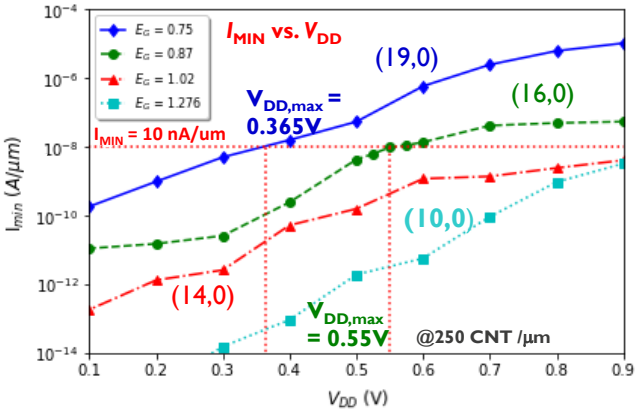

Fig. 2. CNT-FET DFT-NEGF current vs. $V_{DD}$. (a) BTBT-limited leakage current $I_{MIN}$ vs. $V_{DD}$ for the 4 studied zigzag chirality. $L$ = 14 nm. $N_{SD}$ = 2×10$^{20}$ cm$^{-3}$. The current is normalized per device width, assuming a typical CNT density of 250 CNT/ μm [3].

Accurate $E_G$ modelling, coupled with a dissipative QT formalism like NEGF, is essential to assess the fundamentals of CNT transistors, including $I_{MIN}$. To date, however, TB-NEGF models have been used to predict the electrical performance of CNT-FETs [3], [12]. While it might be possible for such models to match experimental $E_G$-$I_{MIN}$ values by using a smaller $d$ than in the experiments and finetuning the TB parameters, the predicted $d$-$I_{MIN}$ relationship is hence quite inaccurate, while this is important for practical technology selection. To the best of our knowledge, the investigation of the fundamentals of CNT-FETs using more accurate HF-DFT NEGF simulations has never been reported.

The relation between $I_{MIN}$ and $V_{DD}$ is investigated using our HF-DFT NEGF model in Fig. 2 for the 4 different zigzag CNT chirality in a n-type $L$ = 14 nm FET assuming a CNT density of 250/μm [3]. Simulated drain-current gate-voltage, $I_D(V_G)$, characteristics of the n-type CNT (16,0) for different $V_{DD}$ are also shown in Fig. 3a. Due to BTBT, in off-state, it is not possible to reduce $I_D$ below $I_{MIN}$, when decreasing $V_G$. $I_{MIN}$ increases with $V_{DD}$ for a given $E_G$.

Direct (ballistic) BTBT from drain (D) to channel (CH) is enabled when D conduction band, $E_C$, is aligned with the CH valence band, $E_V$, i.e., when $V_{DD} \geq V_{DBTBT}$ with:

$$V_{DBTBT} = (E_G - E_{CB}) / q, \quad (1)$$

where $E_{CB}$ is the channel to source barrier required to reach a given off-current level, $I_{OFF}$. Assuming a typical on-current, $I_{ON}$, to $I_{OFF}$ ratio of about 10$^5$ (Fig. 3b):

$$E_{CB} \approx 5* K_B *T*\log(10) \approx 0.3 \text{ eV} \quad (2)$$

In such case, the BTBT current, $I_{BTBT}$, is strong and $I_{MIN}$-($V_{DD}$) typically saturates to its maximum. Additionally, a strong direct $I_{BTBT}$ induces a hole charge pile-up in CH that pins $E_C$, severely degrading SS (Fig. 3a). At lower $V_{DD}$, phonon-assisted BTBT takes over (Fig. 3c), enhanced by the large $E_{OP}$ (≈ 180 meV) in a CNT [12], and the tunnelling probability, hence $I_{MIN}$, decreases with $V_{DD}$ (Fig. 2).

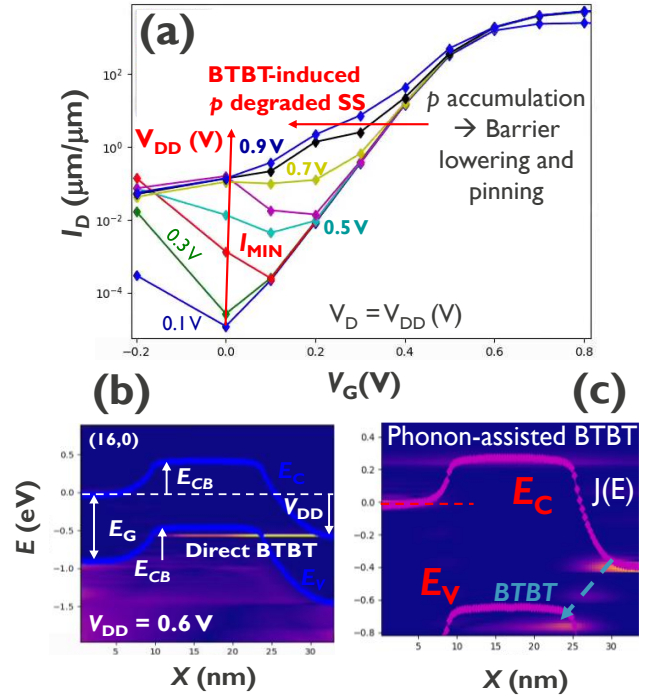

Fig. 3. (a) DFT-NEGF-simulated $I_D(V_G)$ characteristics for the (16,0) CNT-FET vs. $V_{DD}$. $I_{MIN}$ increases with $V_{DD}$. A severe SS degradation due to hole pile-up in the channel is also observed when $V_{DD} > V_{DBTBT}$ = 0.57V (1). HF-DFT-NEGF-simulated $E_C$, $E_V$, and current spectrum, $J(E)$, vs. channel direction, $x$, for the (16,0) CNT-FET at $V_G$ = 0 (OFF-state) with (b) $V_{DD}$ = 0.6V and (c) $V_{DD}$ = 0.4V. In (b), respectively in (c), $V_{DD}$ is larger (smaller) than $V_{DBTBT}$, and the drain $E_C$ overlaps (does not overlap) in energy with the channel $E_V$ and direct (respectively, phonon-assisted) BTBT is present. $E_{CB}$ ≈ 0.3 eV is the source to channel barrier required to get a typical $I_{ON}/I_{OFF}$ ratio of 10$^5$. $L$ = 14 nm. $N_{SD}$ = 2×10$^{20}$ cm$^{-3}$.

For a given $I_{OFF}$, $V_{DD,max}$, the maximum $V_{DD}$ to keep $I_{MIN} \leq I_{OFF}$ can be inferred from the $I_{MIN}(V_{DD})$ curves. For a typical $I_{OFF} = 10$ nA/μm, $V_{DD,max} = 0.35$V for a CNT (19,0), 0.55 V for a (16,0) and larger than 1.1 V for the (14,0) and (10,0) case at $N_{SD} = 2 \times 10^{20}$ cm$^{-3}$ (Fig. 2). The corresponding extracted $I_{ON}$ versus $V_{DD}$ are shown on Fig. 4. For the (19,0) case, the too low $E_G$ induces a low $V_{DD,max}$ and SS degradation, so that $I_{ON}$ is limited to about 1000 μA/μm. For the other studied chirality, however, $I_{ON}$ exceeds $I_{ON,HP-5am}$, the HP drive current for the 2037 IRDS node [1], for $V_{DD} > 0.4$V. The (16,0) CNT-FET reaches the highest $I_{ON} = 4000$ μA/μm at $V_{DD} = 0.5$V, enabling HP logic at $V_{DD} \leq 0.5$ V.

Furthermore, an optimal $N_{SD}$ exists for each chirality. This is shown on Fig. 5 that study the impact of $N_{SD}$ on the $I_{MIN}$-$I_{ON}$ trade-off at $V_{DD} = 0.5$V. Lowering $N_{SD}$ increases the tunnelling distance at D-side, hence, reduces $I_{MIN}$ (Fig. 5a). On the other hand, lowering $N_{SD}$ below a few $10^{20}$ cm$^{-3}$, i.e., the threshold to get $E_C$ degenerately doped, strongly affects $I_{ON}$, as source starvation is observed for lower $N_{SD}$ values (Fig. 5b). In the (16,0) case, $N_{SD} = 2 \times 10^{20}$ cm$^{-3}$ (= 0.25 nm$^{-1}$) is close to optimal and there is not much margin to increase it for the targeted $I_{OFF}$. For the (14,0) and (10,0) cases, on the other hand, increasing $N_{SD}$ to $5 \times 10^{20}$ cm$^{-3}$ (i.e., respectively 0.5 and 0.25 nm$^{-1}$) allows to further increase $I_{ON}$ to 3000 and 2600 μA/μm (Fig. 5b). The (14,0) $I_{ON}$ could reach 5000 and 6000 μA/μm at $V_{DD} = 0.55$ and 0.6V respectively, while maintaining $I_{OFF}$ (Fig. 4).

Finally, at fixed $N_{SD}$ and $V_{DS}$, BTBT, hence, $I_{MIN}$ increases when scaling $L$ due to the increased electric field (Fig. 6). For the (16,0) CNT, $I_{MIN}$ stays below 10 nA/μm down to $L = 9$ nm at $V_{DD} = 0.5$ V and $N_{SD} = 2 \times 10^{20}$ cm$^{-3}$, while the (14,0) (respectively the (10,0)) CNT can maintain the same $I_{MIN}$ down to $L = 6$ (respectively $L \leq 4$) nm, using $N_{SD} = 5 \times 10^{20}$ cm$^{-3}$. For comparison, we also simulated the HfS$_2$ NS $I_{MIN}$ versus $L$. It is similar to that of the CNT (10,0) and stays below 10 nA/μm down to $L = 4$ nm.

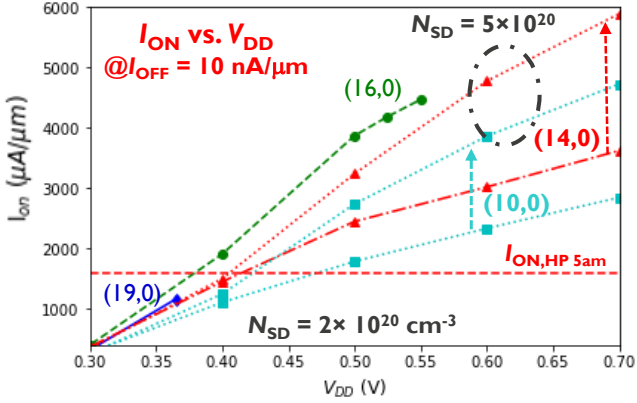

Fig. 4. DFT-NEGF simulated CNT-FET $I_{ON}$ vs. $V_{DD}$ and zigzag chirality at fixed $I_{OFF} = 10$ nA/μm. $L = 14$ nm. $N_{SD} = 2 \times 10^{20}$ cm$^{-3}$ if not stated otherwise. CNT density = 250 CNT/ μm.

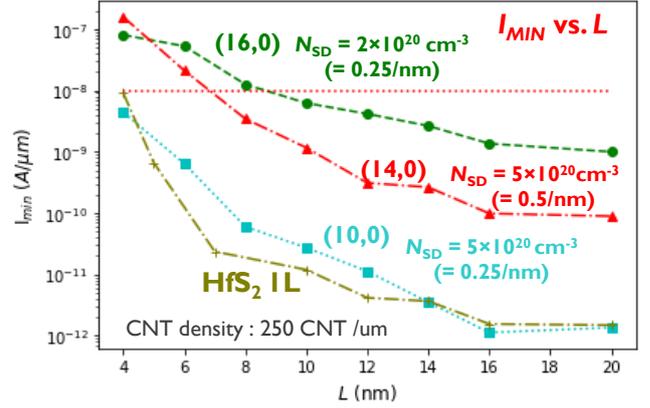

Fig. 6. DFT-NEGF simulated $I_{MIN}$ vs. $L$ at optimal $N_{SD}$ (as indicated on the plot). The HfS$_2$ 1L $I_{MIN}$ ($E_G = 1.202$ eV) is also shown. $V_{DD} = 0.5$V. CNT density = 250 CNT/ μm.

Finally, Fig. 7 benchmarks the $L$ and $V_{DD}$ scaling potential of the CNT-FETs against 2D material MoS$_2$ and HfS$_2$ NS [2] and a 4.4-nm-thin Si NS. In term of short-channel effects, the 2D materials achieve the best SS scaling, followed by the CNTs and then Si (Fig. 7b). The highest $I_{ON}$ is achieved by the CNT-FETs down to $L = 9$ nm ((16,0) is best down to $L = 11$ nm, then (14,0) or (10,0)) that could scale down to $L \approx 8$ nm using $V_{DD}$ in the 0.45-0.5V range. Below $L = 9$ nm, the HfS$_2$ NS offers the best drive and could further scale down to $L = 5$ nm, also using a reduced $V_{DD} \approx 0.5$ V. MoS$_2$ could allow for the smallest $L \approx 4$ nm but using a larger $V_{DD} = 0.6$V. For Si, the $t_{Si} = 4.4$ nm NS scales down to $L \approx 11$ nm at $V_{DD} = 0.6$V.

## IV. CONCLUSIONS

Using Hybrid-Functional DFT coupled with the Non-Equilibrium Green's function formalism, we explored the fundamental physics and performance of scaled CNT-FETs. Using a HF-DFT Hamiltonian in our NEGF model allowed us to significantly improve the accuracy of the $E_G$ vs. $d$ and hence our BTBT, $I_{MIN}$ and $I_{ON}$ predictions, when compared to a TB or standard DFT NEGF simulations. Using our framework, the CNT $I_{MIN}$ and $I_{ON}$ dependency to $d$, $V_{DD}$, doping and $L$ was thoroughly explored.

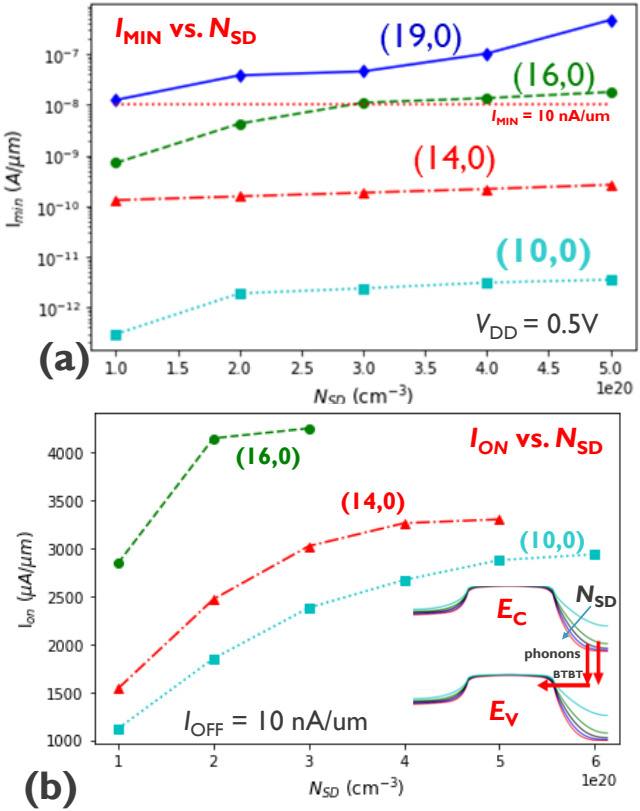

Fig. 5. CNT-FET DFT-NEGF computed current vs. $N_{SD}$ per chirality. (a) $I_{MIN}$ and (b) $I_{ON}$ vs. $N_{SD}$ ($I_{OFF} = 10$ nA/μm) for $L = 14$ nm CNT FETs. The inset shows the (16,0) CNTFET simulated band edges and the related shortening of the BTBT tunneling distance at channel drain-side when increasing $N_{SD}$. $V_{DD} = 0.5$V. CNT density = 250 CNT/ μm.

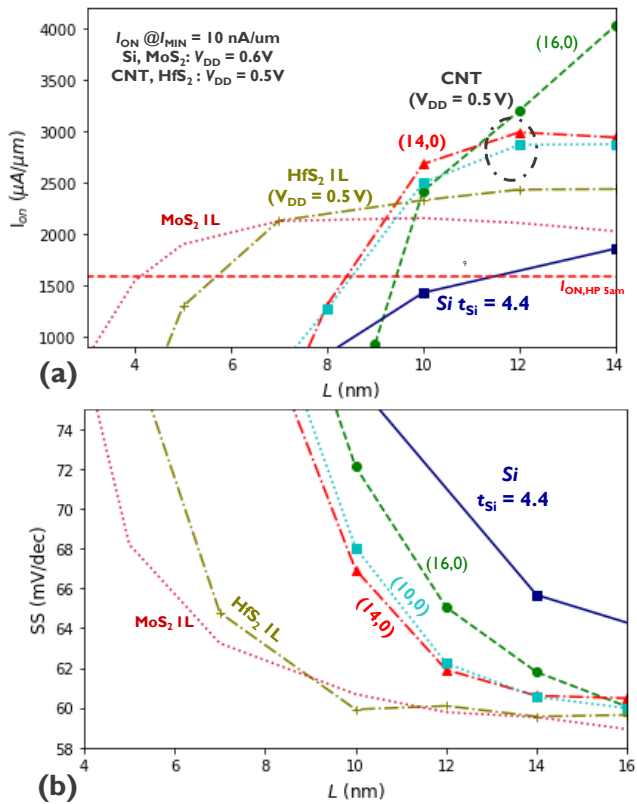

Fig. 7. DFT-NEGF scaling and $V_{DD}$ potential of CNT and 2D-TMDc FETs. (a) $I_{ON}$ vs. $L$ ($I_{OFF}$ = 10 nA/μm) and (b) $SS$ vs. $L$ for optimized Si and MoS$_2$ NS at $V_{DD}$ = 0.6V, as well as high-mobility HfS$_2$ NS and CNT-FETs at $V_{DD}$ = 0.5V. The highest $I_{ON}$ is achieved by the CNT-FETs down to $L$ = 9 nm ((16,0) is best down to $L$ = 11 nm, then (14,0) or (10,0)) that could scale down to $L ≈ 8$ nm using $V_{DD}$ in the 0.45-0.5V range. Below $L$ = 9 nm, the HfS$_2$ NS offers the best drive and could further scale down to $L$ = 5 nm, also using a reduced $V_{DD} ≈ 0.5$ V. The current is normalized per device width.

We further benchmarked the fundamental scaling limits of CNT-FETs against 2D material MoS$_2$ and HfS$_2$ NS and a 4.4-nm-thin Si NS, highlighting the potential for gate length and supply voltage scaling of these different LD material transistors. Optimized CNT-FETs with $d < 1.3$ nm scales better than Si and are promising candidate for high drive current at reduced $V_{DD}$ of 0.5V down to $L$ = 8 nm. Owing to their good-transport properties, they achieve the highest $I_{ON}$ down to $L$ = 9 nm. For lower $L$, however, the 2D materials feature the best SS, and the HfS$_2$ NS offers the best drive with a path for scaling down to $L$ = 5 nm, also using a reduced $V_{DD} ≈ 0.5$ V. Overall, our results highlight a fundamental path towards $L$ = 5 nm, $V_{DD}$ = 0.5V transistors using high-mobility LD materials.

ACKNOWLEDGMENT

The author acknowledges the Imec Industrial Affiliation Program (IIAP) for funding.